# Applicability of Blockchain Technology in Avionics Systems

Harun CELIK and Aysenur SAYIL

*Abstract*— Blockchain technology, within its fast widespread and superiority demonstrated by recent studies, can be also used as an informatic tool for solving various aviation problems. Aviation electronics (avionics) systems stand out as the application area of informatics methods in solving aviation problems or providing different capabilities to aircrafts. Avionics systems are electronic systems used in air and space vehicles for many purposes such as surveillance, navigation and communication. In this study, the applicability of blockchain technology as a new approach in the development of avionics systems is discussed, and in this regard, a method inspired by the previously implemented applications in electronic flight systems is proposed to help evaluate the applicability of this technology in new avionics system designs. The potential of blockchain for solving the problems especially in basic services, communication, navigation and flight management systems; the problem structures for which application of this technology would be a reliable solution; and the superiority and inferiority of its use in avionic systems are explained. A guiding paper is proposed for aviation engineers/experts to make a decision on applying blockchain into avionics systems.

*Authors are with Astronautical Engineering, Erciyes University, Kayseri, Turkiye. Conrresponding author e-mail: haruncelik@erciyes.edu.tr*

# AVİYONİK SİSTEMLERDE BLOK ZİNCİRİ TEKNOLOJİSİ

Harun ÇELİK[1]

Ayşenur SAYIL[2]

## 1. GİRİŞ

Aviyonik kelimesi, havacılık elektroniği (aviation electronics) kelimelerinin birleştirilmesiyle türetilmiştir. Aviyonik sistemler hava araçlarının uçuşunu daha güvenilir, daha düşük maliyetle ve daha kolay gerçekleştirebilmeleri için kullanılan elektronik sistemlerdir. Uçak gibi bir hava aracı üretmenin toplam maliyetinin sivil uçaklar için %35-40, askeri projeler için ise %50'den fazlasını aviyonik sistem geliştirme maliyetleri oluşturmaktadır (Fedosov, Koverninsky, Kan, Volkov, & Solodelov, 2017). Hava araçlarında farklı görevleri yerine getiren birçok elektronik sistem bulunabilmektedir. Bu farklı sistemlerin birbirleriyle entegre olarak çalışmalarını sağlamak son yıllardaki önemli araştırma alanlarından biridir. Bu amaçla oluşturulan entegre modüler aviyonik (Integrated Modular Avionics, IMA) mimarileri daha yüksek düzeyde performans ve sistem kapasitesi sağlamaları, daha fazla donanımın birbirini destekleyerek kullanılabilmesi ve daha düşük bakım-onarım emeği gerektirmesi sayesinde aviyonik sistem maliyetini oldukça düşürmekte ve bu düşüş genel maliyete de önemli katkı sağlamaktadır (Gaska, Watkin, & Chen, 2015). Bunun yanı sıra IMA konsepti sayesinde çok sayıdaki donanımın ve hat üzerinde değiştirilebilir birimlerin (Line Replaceable Unit, LRU) vazifesi, daha az sayıda ve daha merkezi işlem birimleriyle başarılabilmekte, böylece askeri ve ticari hava platformlarında önemli düzeyde ağırlık azaltımı ve bakım tasarrufu sağlanabilmektedir.

Helikopterler de dâhil olmak üzere farklı hava araçlarında herhangi bir yapısal değişiklik gerekmeksizin aynı aviyonik sistemlerin kullanılabilmesi, eklenen yeni sistemlerin eski sistemlerle uyumlu olarak çalışabilmeleri, ortaya çıkacak hataların ek bir teçhizata ihtiyaç duyulmadan sistem içerisinde bulunan fonksiyonlarla tespit edilebilmesi hedeflendiğinden (Collinson, 2011) bu hedefleri gerçekleştirebilen entegre aviyonik sistem kullanımı da gün geçtikçe artmaktadır.

---

[1] Uzay Mühendisliği Bölümü, Erciyes Üniversitesi, ORCID: 0000-0001-5352-3428
[2] Otonom ve Zeki Sistemler Laboratuvarı, Erciyes Üniversitesi, ORCID: 0000-0001-7788-6610

Diğer taraftan aviyonik sistemlerde veriye dayalı karar oluşturma; verinin kullanılabilirliği, tutarlılık ve kesinlik seviyesi gibi çeşitli faktörlerden etkilenmektedir. Pilotsuz uçan hava araçlarının yaygınlaşması havacılık endüstrisinde veri yükünü ciddi boyutta arttırmakta ve büyük boyutlara ulaştığı durumda bu veriler tutarsız olabilmektedir (Alladi, Chamola, Sahu, & Guizani, 2020). Dolayısıyla bu durum, uçağın uçuşa elverişlilik seviyesini, emniyetini ve finansmanını etkileyebilmektedir. Uygulanan yöntemler sistemlerin çok fazla karmaşık hale gelmesine, ortaya çıkan hatalarda sorumlu tarafın kim olduğunun belirlenmesinin zorlaşmasına ve birçok alandaki tasarım mühendislerinin birlikte çalışmalarının zorunlu hale gelmesine neden olduğundan hava araçlarında aviyonik sistem kullanımı kısıtlanabilmektedir.

Blok zinciri teknolojisinde aviyonik sistemleri temel hizmetler konusunda iyileştirebilme, bilgi paylaşımında farklı garantiler sunabilme, güvenliği ve uçuşa elverişlilik seviyesini iyileştirmek için kapsamlı bir veri sağlama kapasitesi bulunmaktadır (Alladi et al., 2020). Blok zinciri, dağıtılmış bir düğüm ağında depolanan, merkezi olmayan, dağıtılmış ve değişmez bir işlem kaydıdır. Hata yapmadan çalışabilen blok zinciri teknolojisi merkezi kayıt tutan teknolojilerle karşılaştırıldığında verilerin şeffaflığını sağlamak gibi önemli avantajlar sunmaktadır (Ahmad et al., 2021). Düğümler veya taraflar arası ağları kriptografik algoritmalarla birleştirerek, bir grup kullanıcının belirli bir durum üzerinde bir anlaşmaya varabilmesini ve bir kontrol otoritesine ihtiyaç duymadan bu anlaşmayı kaydedebilmesini sağlamaktadır (Mehta, Gupta, & Tanwar, 2020). Teknolojik sistemlerin artmasıyla iletişim sinyallerinin birbirine karışmasının yanı sıra, üçüncü kişiler tarafından karıştırılması, verilere dışarıdan müdahale edilerek bozulması, müdahaleler sonucu veri bütünlüğünün sağlanamaması ve hatta hava araçlarının kaçırılması gibi tehditlerin artmasına neden olmaktadır. Blok zinciri teknolojisi, sistemler arasında iş birliğini kolaylaştırmak için güven oluşturabilmektedir.

Bu özellikleri sayesinde blok zinciri teknolojisinin aviyonik sistemlere uygulanması, uçuş operasyonlarının daha güvenli, özerk, esnek ve uygun maliyetli hale getirilmesini sağlayabilir. Günümüzde, hem hava aracı teknolojilerine yapılan yatırımlar hem de hava araçlarına karşı geliştirilen savunma teknolojileri, farklı teknolojiler arasındaki iletişim güvenliğinin sağlanması ihtiyacını önemli kılmaktadır.

Bu bölümde, aviyonik sistemler ile blok zincirinin yapısı ve türleri tanıtıldıktan sonra aviyonik sistemlerde sunulan temel hizmetler, haberleşme, seyrüsefer, uçuş yönetim sistemleri ile bunlarla ilgili problemlerden hangilerinin çözümünde blok zincirinin kullanılabileceği ve hangi yeni kabiliyetlerin kazandırılabileceği irdelenmektedir. Blok zinciri teknolojisini kullanmanın nasıl bir fayda sağlayacağı, bu teknolojiyi kullanmanın etkili olmayacağı problem yapıları, blok zincirinin aviyonik sistemlerde kullanımının faydaları ile sınırları açıklanmaktadır. Bu açıklamalar belirli bir aviyonik sistem veya hava aracı özelinde yapılmayıp temel konular genelinde ele alınmaktadır. Ayrıca örnekler verilirken daha özel yöntemlere de başvurulmaktadır. Örneklerin bu şekilde bir hava aracına özel olduğu durumda da konunun anlaşılması, blok zincirinin farklı hava aracı problemlerine de nasıl uygulanabileceğine dair ufuk açmaktadır. Bu sayede henüz oldukça yeni bir teknoloji olan blok zincirinin tüm aviyonik sitemler açısından nasıl kullanılabileceği veya neden kullanılmaması gerektiği genel hatlarıyla tartışılmakta ve bu alanda çalışanlara rehber bir araştırma sunulmaktadır.

## 2. AVİYONİK SİSTEMLER

Aviyonik sistemler, uçak ve hava taşıtları işletilirken farklı birçok görevin ifasında kullanılan elektronik sistemlerdir. Gün geçtikçe gelişmekte olan bu sistemler arasında uçak yönetim sistemleri, seyrüsefer sistemleri, hava veri sistemleri, uçuş kontrol sistemleri, hava arama ve kurtarma sistemleri, hava trafik kontrol sistemleri, çarpışma önleme sistemleri gibi farklı sistemler bulunmaktadır. Bu sistemler uçak ve hava taşıtlarının güvenli ve etkili bir şekilde uçmasını sağlamaktadır (Kayton & Fried, 1997). Hava aracında bu kabiliyetleri sağlayan aviyonik sistemler, tüm alt sistemleri ile birlikte mimari bir yapı oluşturmaktadır. Bu mimari yapı içerisinde veri iletiminin yapısı, verilerin işlenme şekli, veri yolları, kullanılan sensörler ile mimari yapının kapladığı yer ve ağırlığı; aviyonik sistemin performansını, kapasitesini ve maliyetini doğrudan etkilemektedir. Sistem mimarisini etkileyen bu temel işlevlerin yanında hava aracında uygulanan son teknolojilerle mimariye yeni işlevsellikler de eklenebilmektedir.

Bir aviyonik sistemde, temel işlevlerin her birinde sağlanacak ilerlemeler aviyonik açıdan önemli avantajlar sunabilir. Bunu sağlamak için son zamanlarda entegre modüler aviyonik mimarilerin büyük önem kazandıkları görülmektedir (P. Han, Zhai, & Zhang, 2020; Wang & Niu, 2018). Bu mimariler sayesinde aynı aviyonik sistemin farklı hava araçları için değişiklik gerektirmeden uygulanabilmesi, eklenen yeni sistemlerin eski sistemlerle uyumlu olarak çalışabilmesi, ortaya çıkan arızaların ek bir teçhizata ihtiyaç duyulmadan sistemin kendi içerisinde bulunan fonksiyonlarla tespit edilebilmesi hedeflenmektedir. Bu bağlamda daha hızlı ve gelişmiş veri yolları oluşturulmakta veya Şekil 1'de gösterildiği gibi klasik bağlantı elemanları olan LRU'lar modüler olarak tasarlanmaktadırlar (Collinson, 2011). Ancak bu yöntemler sistemlerin çok daha karmaşık hale gelmesine, ortaya çıkan hatalarda sorumlu tarafın kim olduğunun belirlenmesinin zorlaşmasına ve birçok alandaki tasarım mühendislerinin birlikte çalışmalarını zorunlu kılmaya neden olduğundan hava aracında uygulanabilirliği zor ve kısıtlı olmaktadır.

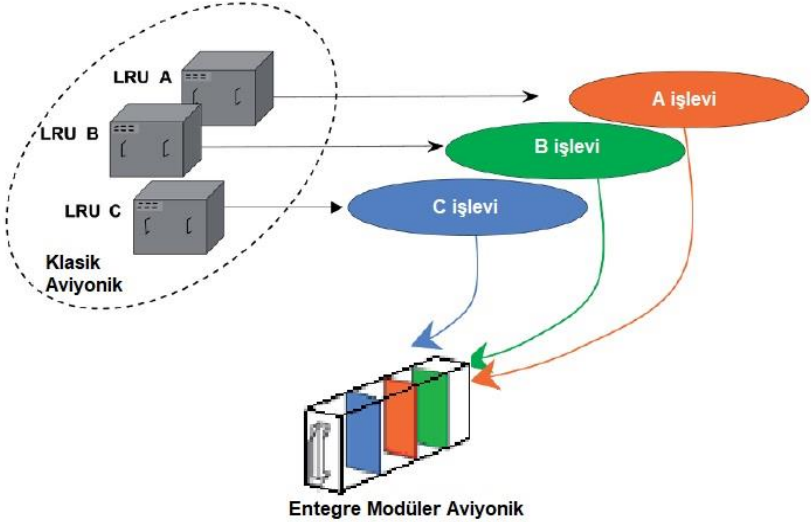

*Şekil 1. Aviyonik sistemlerin entegre edilmesi*

Teknolojinin sürekli ilerlemesine ve nesneler/teçhizatlar arası iletişimin artmasına paralel olarak modern hava araçlarında da aviyonik sistemlerin artık sınırlı etkileşimli bir sistem olarak kullanılmasından öte diğer aviyonik sistemlerle eşgüdümlü olarak birlikte ve çok daha verimli çalışması için birçok araştırma yapılmaktadır (Chen, Du, & Han, 2016; S. Han & Jin, 2014; Miller, Cofer, Sha, Meseguer, & Al-Nayeem, 2009). Bu araştırmaların ana hedefinde hava araçlarında bulunan elektronik sistemlerin uyum içinde ve birlikte çalışabilecekleri sistemler tasarlamak ve son teknolojilerin uygulandığı yeni mimariler oluşturmak bulunmaktadır. Dolayısıyla aviyonik sistemler sınırlı etkileşimli bir sistem olmaktan öte IMA'da olduğu gibi tüm aviyonik sistemlerin birbirleriyle eşgüdümlü çalıştıkları, yeri geldiğinde iki sistem arasında bir bağlantı olmasa dahi başka bir sistemi aracı yaparak iletişim kurabildikleri, uçuş yönetim sistemi gibi aviyonik sistemlerin yönetim fonksiyonunun çok daha geliştiği bir konuma gelmektedir.

Hava aracı aviyonik sistemleri arasındaki ARINC veya MIL-STD ile belirlenen standartlardaki dahili veri akışı ile uçuş veri kayıt (Flight Data Recorder, FDR) sistemlerinin güvenliği kadar dış aviyonik sistemlerle (hava trafik kontrolü, çarpışma önleme sistemleri gibi) olan bağlantılar da modern ihtiyaçları karşılayacak şekilde yeni yöntemlerin uygulanmalarını zorunlu kılmaktadır. Nitekim gün geçtikçe artan hava aracı trafiği de, hava aracının kendi güvenli uçuşuyla birlikte diğer hava araçları ve yer istasyonları ile sağlıklı bir etkileşim kurmayı gerektirmektedir. Hava aracı sayısı arttıkça trafikte çıkabilecek sorunların risk oranı da artmaktadır. Tüm bu aviyonik sistemlerin gelişiminde blok zinciri teknolojisinin önemli bir rol alabileceği görülmektedir.

## 3. BLOK ZİNCİRİ TEKNOLOJİSİ

Blok zinciri, tüm verilerin şifreleme yöntemleri kullanılarak merkezi olmayan bir yapıdaki veri tabanında tutulduğu, tutulan kayıtların da değiştirilmesinin çok zor olduğu bir teknolojidir (Juma, Shaalan, & Kamel, 2019). Bu yapıda, her işlem sistemde bloklar halinde tutulmaktadır (Mehta et al., 2020). Karma şifreleme algoritmasıyla oluşturulan her blok kendinden önce oluşturulmuş bloğun bilgilerini tutan şifrelenmiş bir değer de taşımaktadır (Zhu, Zheng, & Wong, 2019). Bloklar bu şifrelenmiş değerlerle birbirine bağlanmakta, her blok önceki bloğun özet (hash) değerine kendi blok başlığında yer vermektedir. Bu sayede, blok zincirinde bulunan herhangi bir özet değerinin değişmesi durumunda ağda bulunan tüm değerlerin değişmesi gerekmektedir (Mehta et al., 2020; Zhu et al., 2019). Birbirini tetikleyen bu durum nedeniyle tüm blok değerlerinin yeniden hesaplaması gerekecektir. Çok büyük boyuttaki yeniden hesaplama işlemi de sistem üzerinde büyük yük oluşturduğundan gerçekleştirilmesi çok zordur. Bu da blok zincirin değiştirilemez olmasını sağlamaktadır (Ahmad et al., 2021). Her bir bloğun bir önceki bloğa ait özet değerini içermesi vasıtasıyla kaydedilmiş işlemlerde kötü niyetli değişiklikler yapılması önlenmektedir (Akleylek & Seyhan, 2018).

Merkezi kayıt tutma biçimleriyle karşılaştırıldığında blok zinciri hem verilerin tam şeffaflığını sağlamakta hem de kötü niyetli girişimleri engellemektedir. Blok zinciri teknolojisi bir yandan taraflar için anonimlik ve güvenlik sağlarken diğer yandan da onay için bir aracıya veya üçüncü bir kişiye olan ihtiyacı ortadan kaldırmaktadır. Örneğin bir veri tabanına yalnızca bir kişi sahip olduğunda yanlışlıkla veya kasıtlı olarak hata yapılma olasılığı bulunmaktadır. Fakat sistemdeki herkes bu veri tabanının bir kopyasına sahipse, yani veri tabanı tek değilse, hile yapmak zorlaşmaktadır. Bu özellikleri barındıran blok zinciri teknolojisinin güvenli, esnek, değiştirilemez, merkezi olmayan ve şeffaf bir veri yapısına sahip olması bu teknolojisinin farklı birçok alanda tercih edilmesine neden olmuştur (Alladi et al., 2020; Zhu et al., 2019). Blok zincirinin bu üstünlüklere nasıl sahip olduğunun anlaşılabilmesi için bu teknolojinin dayandığı temellerin anlaşılması gerekir.

### 3.1. Blok Zincirinin Yapısı

Blok zinciri yapısı itibariyle veri, ağ, mutabakat, teşvik, sözleşme ve uygulama katmanlarından oluşmaktadır. Veri katmanı, zaman damgalı veri bloklarını içeren blok zinciri mimarisinin en alt katmanıdır (Alladi et al., 2020). Bloklar verilerin saklandığı yapılardır. Blok zincirinde belirli bir işlem depolama kapasitesine sahip bir blok, belirli bir zaman diliminde ağa eklenmek amacıyla oluşturulmaktadır (Güven, 2020). Her bir veri bloğu, blok gövdesi ve blok başlığından oluşmaktadır. Bloğun başlık kısmı ağın versiyonunu, bir önceki bloğa ait özet değerini, Nonce sayısını, zorluk hedefini, oluşturulma zamanını, Merkle kökü değerini barındırmaktadır.

Blok gövdesinde ise veriler tutulmakta ve içerisinde doğrulaması yapılmış işlemler bulunmaktadır. Blok zincirde blok içerisindeki tüm işlemler önce tek tek, sonra da ikişer ikişer özet değeri fonksiyonuna sokulmakta, en son elde edilen özet değeri de Merkle kökü olmaktadır (Aslan & Kasapbaşı). Blok zincirinde her bir blok kendinden önceki bloğun özet değerini içermekte, böylece blok içerisinde saklanan işlemlerin değiştirilmesi ya da silinmesini zorlaştırılmaktadır (Aslan, 2022). Bu sayede herhangi bir saldırganın, blok zincir ağında hedef aldığı blokların içeriklerini değiştirebilmesi için hedef bloklarla birlikte bu blokları takip eden diğer tüm blokları da değiştirmesi gerekmektedir.

Yapıda yer alan kriptografik özet değeri fonksiyonu, matematiksel fonksiyonlar kullanarak değişken boyutlu işlem girdisinden sabit boyutlu eşsiz değer üreten işlemdir. Herhangi bir veri girdisinde ufak bir değişiklik dahi yapılması tamamen farklı bir özet değerin üretilmesine sebep olmaktadır. Blok içindeki ağın versiyonu, önceki bloğun özet değeri, zorluk hedefi, işlem adedi gibi veriler sabit olduğunda farklı özet değerleri üretebilmek maksadıyla rastgele oluşturulan bir sayı yani Nonce değeri fonksiyona eklenir. Nonce sayı değeri zorluk hedefine göre belirlenen rastgele bir sayı olarak tanımlanabilir. Nonce değerini tek başına bulmaya çabalayan bir tarafın veya düğümün verdiği uğraş bir yıl kadar uzun zaman gerektirebilmekteyken, ağdaki tüm düğümlerin eş zamanlı verdikleri uğraş Nonce değerinin birkaç dakika kadar kısa bir zamanda bulunabilmesini sağlamaktadır. Düğümün işlem gücünün artması veya daha fazla düğümün sisteme eklenmesi ile blokların üretilmesi için gerekli süre oldukça kısalabilmektedir.

Birinin hızlıca Nonce değeri bularak blokları seri bir şekilde eklemesini engellemek için dinamik olarak değişebilen bir zorluk hedefi belirlenmektedir. Şayet ağdaki işlem gücü artar ve blokların eklenmesi belirlenen süreden daha kısa sürerse sistem zorluk seviyesini otomatik olarak artırmakta, ağdaki işlem gücünün azalması ve dolayısıyla blok eklenme süresinin belirlenen süreden daha uzun sürmesi durumunda da sistem zorluk seviyesini azaltmaktadır. Belirlenen zorluk seviyesine göre özet değerinin kaç hanesinin sıfırla başlayacağına yani boyutuna karar verilmektedir.

Blok zinciri teknolojisinin ağ katmanı ile de işlemler dağıtılmakta, iletilmekte ve doğrulanmaktadır. Merkezi mimariden farklı olarak genellikle tarafların eşit haklara sahip olduğu (peer-to-peer, P2P) ağlar olarak tasarlanan blok zincirlerinde bir işlem, oluşturulduktan sonra doğrulama için komşu düğümlere yayınlanmaktadır (Alladi et al., 2020). Bu süreç, önceden tanımlanmış özelliklere göre gerçekleştirilmektedir. İşlem doğrulandıktan sonra diğer düğümlere gönderilmekte ve düğümlerin reddetmesi durumunda işlem iptal edilmektedir. Bu da her düğümde yalnızca geçerli işlemlerin kaydedilmesini sağlamaktadır.

Blok zincirinde güvenlik amacıyla asimetrik bir kriptografi mekanizması kullanılmaktadır. Bu sayede yalnızca özel anahtar sahibinin şifreyi çözebileceği şekilde veriler şifrelenebilmekte veya verileri özel bir anahtarla imzalayarak onların güvenilir bir kaynaktan geldiği kanıtlanabilmektedir. Blok zincirinde bu mekanizma kullanılırken bahsedilen bu iki temel kriptografik fonksiyon veri doğrulama ve veri sahipliğinin kanıtı aşamaları olarak adlandırılmaktadır (Kamel Boulos, Wilson, & Clauson, 2018). Bu mekanizmaya bir örnek insansız hava araçları (İHA'lar) üzerinden Şekil 2 ile sunulabilir.

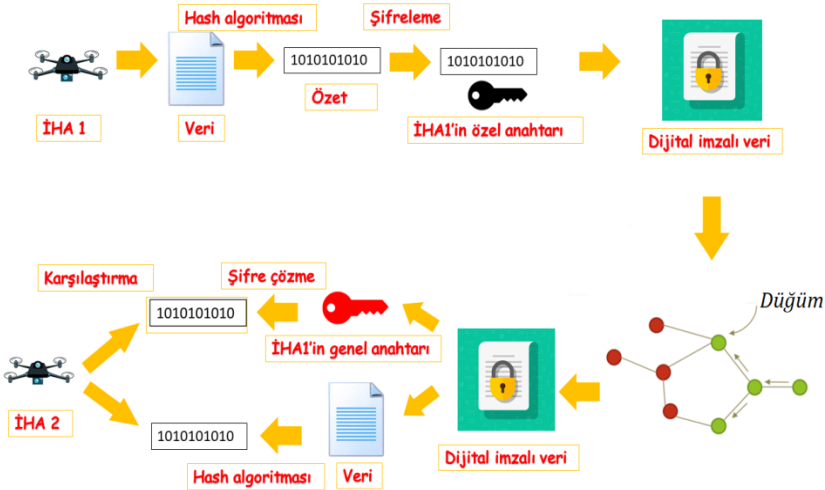

***Şekil 2: Blok zinciri ağında veri doğrulanması ve veri sahipliğinin kanıtlanması***

Burada İHA 1, İHA 2'ye bir veri göndermek istediğinde, önce verinin özeti oluşturulmakta sonrasında özet sadece İHA 1'de bulunan özel bir anahtarla kilitlenmektedir. Özel anahtar önemlidir ve kimseyle paylaşılmamaktadır. Veri sahipliğinin kanıtı için İHA 1 bu özel anahtarını kullanarak dijital imza oluşturabilmekte ve kendisinin imza sahibi olduğunu kanıtlayabilmektedir. İHA 1 bu dijital imzayı, açık anahtar ve iletilecek belgeyle beraber doğrulama yapılması için tüm ağa yayar. Bu düğüm dijital imza ve genel anahtarı doğrulama algoritmasına so-kulur. Şifrelenmemiş özet, alınan belgenin oluşturulmuş bir özetiyle karşılaştırılır. Doğrulama işleminin başarılı olarak yapılmış olması için bu karşılaştırma sonucunun eşleşmesi gerekir (Kamel Boulos et al., 2018). Yani İHA 1'in gönderdiği veri başka birinin özel anahtarı ile imzalanmışsa, İHA 1'in genel anahtarı ile doğrulanmayacaktır, sadece İHA 1'in özel anahtarıyla imzalanmışsa doğrulanacaktır. Doğrulama işlemi gerçekleştirildikten sonra işlem bloğa eklenir. Sonrasında madencilik süreci başlar. Yani taraflar matematiksel bir problemi çözmek için yarışmaya başlarlar. Kazanan zincire yeni blok ekleme hakkı elde eder. Eklenen yeni blok ile birlikte her bir düğüm sahip olduğu özeti günceller. İHA 1'den İHA 2'ye veri aktarımı gerçekleşir.

Mutabakat katmanı, blok zinciri ağına katılan güvenilmeyen taraflar arasında bir fikir birliğine varmak için gerekli olan farklı fikir birliği algoritmalarından oluşmaktadır (Kumar, 2021). Katılımcılar arasındaki fikir birliği, merkezi bir varlığa olan ihtiyaçtan kaçınmanın anahtarıdır. Bu nedenle, blok zinciri ilkesinin temelini oluşturmaktadır. Blok zinciri tarafları arasında fikir birliği sağlamak için bazı protokollere ihtiyaç duymaktadır. Pratik Bizans Hata Toleransı (PBFT), İş Kanıtı (PoW) ve Hisse Kanıtı (PoS), Yetkilendirilmiş Hisse Kanıtı (DPoS) fikir birliği algoritmalarının başlıcalarıdır. İş kanıtı (Proof of Work) algoritmasının en düşük hata payına sahip olmasına rağmen enerji verimliliğinin düşük olduğu görülmektedir.

Teşvik katmanı ise düğümler için karşılıklı fayda sağlayan bir plan oluşturmaktadır (Alladi et al., 2020). Sözleşme katmanıyla blok zinciri programlanabilir hale gelmekte ve blok zinciri üzerinde karmaşık işlemlerin gerçekleştirilebileceği farklı komut dosyalarının, akıllı sözleşmelerin ve algoritmaların dahil edilmesini sağlamaktadır (Alladi et al., 2020). Uygulama katmanı havacılık, uzay veya finans, nesnelerin interneti gibi çeşitli pratik alanlardaki uygulamalardan oluşan blok zincirin en üst katmanıdır.

Blok zinciri teknolojisinin önemli uygulamalarından biri de akıllı sözleşmelerdir. Akıllı sözleşme, katılımcılar arasında yapılması düşünülen bir anlaşmanın sözleşmesinin tüm maddelerinin bilgisayarla kodlanması ve bu sayede denetimi dijital ortamda yapılabilen sözleşmeleri mümkün kılmaktadır. Bir akıllı sözleşme, katılımcılar arasında güven oluşturan, önceden belirlenmiş şartlar çerçevesinde yürüyen bir yazılımdır. Akıllı sözleşme, kullanıcıların arasında anlaşmalar yapmasına blok zinciri ağı ile olanak tanımaktadır. Bu sözleşmeler aşağıdan yukarıya inşa edilmiş olan bir Merkle karma ağacı şeklinde tasarlanır. Veri koleksiyonlarını içeren belirli bir blok zincir adresinde kaydedilmiş ve önceden tanımlanmış talimatlar bulundurur. Akıllı sözleşmeler güveni, verimliliği, doğruluğu ve özerkliği arttırmaktadır. Bu özellikleri tarafların üzerinde mutabık kaldığı sözleşmelere noter gibi herhangi bir üçüncü taraf olmadan zor ve uzun işlem sü-recini kolaylaştırması ve hızlandırması için kullanılmaktadır. Noter gibi onay mekanizmalarının yazılım almaktadır (Oğuzhan & Kiani, 2018). Bundan dolayı, blok zinciri ve akıllı sözleşmelerin birlikte kullanımının merkezi olmayan aviyonik sistemlerde uygulanması, güvenli veri paylaşımı ve erişim yetkisi sağlama kapasitesi sunmaktadır.

### 3.2. Blok Zincirinin Türleri

Blok zincirlerini herkese açık olanlar, özel olanlar ve iş birlikteliğinde kullanılanlar olmak üzere üç ana sınıfta toplamak mümkündür. Bunlardan herkese açık veya genel blok zincirinde tüm kayıtları herkes görebilmekte ve fikir birliği sürecine katılabilmektedir. Açık blok zincirlerinde katılımcı sayısı çok yüksek olduğu için diğer iki türe kıyasla en yüksek değişmezliğe, fakat daha düşük verimliliğe sahiptir (Alladi et al., 2020). Kripto para teknolojileri tamamen dağıtık yapıdaki herkese açık blok zinciri ağını kullanmaktadır. Herkese açık olmakla ağ, güvenilirliğini yitirmemektedir. Nitekim ağdaki herkesin yaptığı işlemler açıkça görülmekte ancak kişisel bilgiler ise bilinmemektedir (Aslan, 2022).

Özel blok zincirinde sadece bir kurumdan ya da müşteriden gelen düğümlerin ağa ve fikir birliği sürecine katılmasına izin verilmektedir, yani izinli bir fikir birliği sürecine sahiptir. Tamamen tek bir organizasyonun kontrolü altında olduğu için merkezi bir ağ olarak da kabul edilmektedir. Bu tür ağlar yüksek verimliliğe sahiptir, ancak daha az sayıda katılımcı olması nedeniyle veriler halka açık blok zincirlerinden daha kolay değiştirilebilmektedir. İş birliğinin veya konsorsiyum blok zincirinde ise yalnızca birkaç seçilmiş kuruluşun katılmasına izin verilmektedir yani izinli bir fikir birliği sürecine sahiptir. Bu nedenle, kısmen merkezi olmayan bir sistemdir. Aynı zamanda yüksek verimliliğe sahiptir, değiştirilebilme veya karıştırılabilme yönünden herkese açık blok zincirleri kadar olmasa da özel blok zincirlerinden daha güvenilirdir (Alladi et al., 2020). Blok zinciri türlerine kullanılacağı sistemlere ve kullanım amaçlarına göre karar verilmesi gerekmektedir.

### 4. AVİYONİK SİSTEMLERDE BLOK ZİNCİRİNE OLAN İHTİYAÇ

Aviyonik sistemler arasında merkezi ve öncü bir konumu olan uçuş yönetim sistemi (flight management system, FMS) hava aracında temel denetim birimi olarak işlev görmektedir. Modern hava araçlarında nesneler arası iletişimin artması FMS'yi sadece sınırlı etkileşimli bir sistem olmaktan çıkarmıştır. Böylece, FMS diğer aviyonik sistemlerle verimli bir şekilde eşgüdümlü çalışabilmektedir (Chen et al., 2016; S. Han & Jin, 2014). Gelişmiş bu FMS'lerde blok zinciri teknolojisi de hava araçlarında yolculuk ve uçuş güvenliğini artırmaya yardımcı olacak şekilde kullanılabilmektedir. Blok zinciri teknolojisine sahip FMS sistemleri sayesinde uçakların bakım ve onarım geçmişi izlenerek uçakların daha güvenli bir şekilde uçması sağlanmaktadır. Ayrıca bu tür FMS'ler ile hava trafiği yönetim sistemleri aracılığıyla, uçuş güzergâhlarının tasarımı, uçuş izleme sistemleri ve uçuş kontrol sistemleri gibi birçok farklı sistem eşgüdümlü çalışabilmektedir (Clementi, Larrieu, Lochin, Kaafar, & Asghar, 2019).

Sivil havacılıkta gelişen aviyonik sistemlerden biri de Otomatik Bağımlı Gözetleme Yayını (ADS-B) teknolojisidir, bazı ülkelerde bu teknolojinin kullanımı zorunlu hale getirilmiştir. ADS-B uçakların pozisyonunu, hızını ve diğer yolculuk bilgilerini gerçek zamanlı olarak iletmek için kullanılan bir teknolojidir. Bu teknoloji, GPS veya GLONASS gibi konum belirleme sistemleri kullanarak uçağın pozisyonunu belirlemekte ve bu bilgiyi hava trafik kontrolörüne ile diğer uçaklara iletmektedir. Bu sayede, hava trafik kontrolörü daha etkili bir şekilde uçakların yönünü ve hızını yönetebilmekte ve uçaklar arasındaki potansiyel çarpışma riskini azaltabilmektedir.

Öte yandan bu teknoloji aynı zamanda güvenlik riskleri de taşımaktadır. Örneğin, gönderilen bilginin sahte olması ve değiştirilmesi, bilginin yanlış yönlendirilmesi veya gönderimin engellenmesi, gecikmesi gibi problemler mevcuttur. Bu nedenle, ADS-B sistemi kullanılırken güvenliğin sağlanması için önlemler alınmalıdır. Uçakların konum bilgisinin blok zinciri ile saklanıp kaydedilmesi ile sistemin güvenliğinin arttırılması mümkündür (Gaska et al., 2015). Seyrüsefer iletişiminde gecikmenin azaltılıp, az gecikmeyle iletişim güvenliğini sağlamak için akıllı sözleşmeli bir blok zinciri kullanılabilmektedir (Blasch, Xu, Chen, Chen, & Shen, 2019).

Ülkemizdeki kayıtlı sivil hava araçlarının sayısı son yirmi yılda yüzlü rakamlardan atmış binin üzerine çıkmıştır. Başta insansız hava araçları olmak üzere sayı her geçen gün daha fazla artmaktadır. Aviyonik sistemlerin kullanıldığı hava araçlarının önemli bir kısmını da İHA'lar oluşturmaktadır. İHA'lar uçuş görevlerinde yük taşıyabilen, uzaktan kumanda istasyonları tarafından veya otonom bir şekilde yönlendirilen robotik araç sınıfı olarak tanımlanmaktadır. 2030 yılına kadar Japonya topraklarında günlük olarak yaklaşık 210 000 adet İHA ile devriye gezmeyi planlamaktadır (Yaguchi & Wakazono, 2021). İHA'ların birçok alanda kullanımı verimlilikte artış sağlarken, aynı zamanda güvenilmeyen üçüncü şahıslardan gelen siber tehditlere ve saldırılara da kapı açmaktadır.

Hava aracı sayısı arttıkça aralarındaki iletişim, güvenlikleri, hava veri güvenliği, veri depolama ve yönetimi gibi hava aracı ağlarındaki çeşitli sorunlar da artmaktadır. Hava araçları kaybolmaya, yok edilmeye veya kaçırılmaya karşı savunmasız kalmaktadır (Alladi et al., 2020; García-Magariño, Lacuesta, Rajarajan, & Lloret, 2019). Farklı görevleri birlikte icra ederken grup üyeleri arasındaki iletişim dış müdahalelere açık hale gelebilmektedir (Millard, Timmis, & Winfield, 2013). Grup veya sürü hava aracı sistemleri de dahil olmak üzere herhangi bir ortamda güvenlik, temel olarak verinin gizliliği ve doğruluğu, varlık kimlik doğrulaması ve veri kaynağı kimlik doğrulaması gibi temel hizmetlerin sağlanması ile ilgilidir. Bu tip güvenlik sorunlarına pratik çözümler bulma konusunda ise sıkıntı yaşanmaktadır (Higgins, Tomlinson, & Martin, 2009). Blok zinciri teknolojisi bu sorunlara ve zorluklara çözüm sunacak bir yol olarak görülmektedir. Blok zinciri şifreleme şemasında, açık anahtar ve dijital imza kriptografisi gibi teknikler, yalnızca güvenli olmayan ve paylaşılan kanalları kullanarak işlem yapmanın değil, aynı zamanda bir ağdaki belirli araçların kimliğini kanıtlamanın da yöntemleridir (Ferrer, 2016). Blok zinciri teknolojisi hem hava araçlarına güvenilir bir düğümler arası iletişim kanalı sağlama hem de farklı tehditlerin, güvenlik açıklarının ve saldırıların üstesinden gelme potansiyeline sahiptir.

### 4.1. Uçuş Güvenliği

Uçuş yönetim sistemi, uçuş bilgilerinin aktarılmasını sağlayan haberleşme sistemleri ile hava trafik yönetimi uçuş güvenliği açısından en önemli aviyonik sistemleri oluşturmaktadır. Uçuş yönetim sistemi (FMS) gibi aviyonik sistemler, yakıt ve zaman tasarrufu açısından en verimli uçuşun hesaplanması, uçuş planının oluşturulması ve pilotların navigasyon sistemi ile uçuş hazırlığında işlerini kolaylaştırmak için de kullanılmaktadır (Yaguchi & Wakazono, 2021). Uçuş ile ilgili verilerin paylaşıldığı önemli sistemler arasında ADS-B (Automatic Dependent Surveillance Broadcast) ve ACAS (Airborne Collision Avoidance System) sayılabilir. Hava trafik yönetimi ise ATM (Air Traffic Management) sistemi tarafından gerçekleştirilmektedir. Bu sistemlerin herhangi birindeki güvenlik açığı uçuş güvenliğini doğrudan tehdit etmektedir.

Uçaklarda ve helikopterlerde ATM sisteminin taşıdığı önemin benzerini insansız hava araçlarında da yer kontrol istasyonu ile olan iletişim taşımaktadır. Kablosuz ağların yoğun kullanımı ve düşük ağ kapasitesi, ulaşılabilen alanın alanının dar olması ve önleyici önlemlerin alınmaması gibi faktörler kablosuz sinyal yoğunluğuna sebebiyet vermektedir. Bu yoğunluğun saldırı amacıyla oluşturulduğu ve bu nedenle oluşan güvenlik zafiyeti sonucunda bir mühendisin hayatını kaybettiği bir kaza da mevcuttur (Kuzmin & Znak, 2018). Dolayısıyla uçuş kontrol ve seyrüsefer sistemlerinin yer kontrol istasyonundan gelen komutlara bağlı olduğu durumlarda kablosuz sinyallerdeki yoğunluk ölümcül olabilmektedir.

Güvenliğin sağlanması gereken yerlerden biri de hava aracı bakım kayıtlarıdır. İşletilen hava araçlarının bakım süreçleri uçak bakım kayıt defterleri ile belgelenmektedir. Hava aracı havalanmak için mükemmel durumda olsa dahi uygun bakımın tamamlandığını gösteren defter olmadan uçuşa elverişli kabul edilmez. Ancak, kayıtların doğru tutulması, bir uçağın uçuşa elverişliliğinin sağlanmasında kritik bir prosedürdür. SHGM'nin ilgili yönetmeliğine göre bu kayıt defterleri sınıfına göre belirli bir süre saklanmak da zorundadır. Saklama yöntemleri (fiziksel veya elektronik ortamda saklama) hava aracı sahiplerinin tercihine bırakılmakla birlikte genellikle ilgili uçakta veya havayollarının mülkiyetinde fiziksel bir defter olarak saklanmaktadır (Aleshi, Seker, & Babiceanu, 2019). Kayıtların hem tutulmasında hem de saklanmasında oluşabilecek güvenlik tehditleri hava aracı, yolcu ve işletici açısından farklı kötü sonuçlar doğurabilir. Dağıtık bir depolama hizmetine sahip blok zincir teknolojisi, ilgili uçağın servis, bakım, sertifika gibi önemli kayıtlarını süresiz, şeffaf ve güvenli bir şekilde saklama imkanı sunabilmektedir.

### 4.2. Milli Güvenlik Riskleri ve Mahremiyet

Havaalanları, askeri alanlar, kamu kurumları gibi özel alanlar üzerinde uçan veya bir yükü bir noktadan diğerine taşımak için kullanılan hava araçları kurum ve toplum için mahremiyet, gizlilik gibi ciddi sorunlar teşkil edebilir (Blank, Kirrane, & Spiekermann, 2018). Özel mülklere girmeleri durumunda hava araçları kaçırılarak da insanların özel mülklerinin, ticari alanlarının veya askeri alanların mahremiyeti ihlal edilebilir. Bu ihlaller istihbarat veya şantaj amaçlı da kullanılabilir. Bu nedenle hava araçlarıyla elde edilen verilerin iyi bir şekilde korunmasına ve gizli tutulabilmesine ihtiyaç duyulmaktadır.

Mahremiyetin korunması ve gizliliğin temini için hava aracı uçuş planlarının bunları dikkate alarak oluşturulması gerekebilir. Özellikle İHA tabanlı hizmetlerde özel, genel ve endüstriyel uçuşlar sırasında mahremiyetin korunması gerekir (Rahman, Khalil, & Atiquzzaman, 2021). Bu da uçuş planları oluşturulurken gerekli olan irtifa, hareket yönü, geçilecek noktaların sayısı, azami ve asgari sürat gibi bilgilerin yanında mahrem alanların konum bilgilerinin de hesaplamaya dahil edilmesini gerektirir. Bu şekilde oluşturulan uçuş planlarında uçuşun gerçekleştirilmesi daha maliyetli ve uzun olsa da bu uçuş planlarının değiştirilememesi önemlidir. Ayrıca tüm hava araçlarının uçuş kayıtları alınarak daha sonraki incelemeler için saklanması ve bunlara kimsenin geriye dönük olarak müdahalede bulunamaması için değiştirilmesi zor olan bir şekilde kayıt yapan sistemlere ihtiyaç durulmaktadır.

### 4.3. Veri Güvenliği

Hava araçlarının hem birbirleri hem de yerle olan iletişimlerinin güvenliği de büyük önem arz etmektedir. Hava araçlarında veri üreten Sonar darbe mesafe sensörü, ışık darbesi mesafe sensörleri, uçuş süresi sensörleri, termal sensörler, kimyasal sensörler ve yönelim gibi birçok sensör mevcuttur ve bunların ürettiği verilerin paylaşılması gerekir (Bera, Chattaraj, & Das, 2020). Hava trafiğinin yoğun olduğu bölgelerde her bir sensör verisinin paylaşılması sayesinde bu sensörlere sahip olmayan hava araçları da verileri kendileri üretmek zorunda kalmadan elde edebilirler. Bu gibi amaçları gerçekleştirmek için son yıllarda yapılan nes-nelerin interneti (internet of things, IoT) çalışmaları da hız kazanmıştır. Nesnelerin interneti ağına bağlı cihaz sayısını artması, sinyallerin uzaktan karıştırılması, mesajların silinmesi veya değiştirilmesi gibi güvenlik saldırılarıyla karşılaşılmaktadır. Bu ağa dâhil edilmiş bir blok zincirine sahip olmak, süreci merkezi olmayan ve özerk hale getirecektir.

Klasik hava araçları ile birlikte haberleşme, ölçüm, sensör, hesaplama sistemleri de genellikle az enerji harcayan, düşük hesaplama gücüne sahip bir yapıdadırlar. Hava araçlarında sistem ve hesaplama karmaşasından mümkün oldukça kaçınılır. Ancak düşük işlem kapasitelerinden dolayı iletilen verileri kötü niyetli girişimlerden korumak için karmaşık şifreleme yöntemleri de çalıştıramazlar. Araçlar arasındaki iletişimde güven eksikliği, özellikle yer kontrol istasyonuyla da beraber gerçek zamanlı çalışmaları gerektiğinde, yapılan veri alışverişi olası saldırılara daha açık hale gelmektedir. İnsansı hava araçları aralarındaki ağ daha savunmasız ve dinamik olduğu için yanlış yönlendirmeler yapılabilmektedir (Mehta et al., 2020). Bu tür saldırıların üstesinden gelmek için yeni yöntemlerin uygulanması gerekmektedir.

### 4.4. Sürü Güvenliği

Son yıllarda farklı türdeki hava araçlarının iş birliği içerisinde gerçekleştirdikleri ve özellikle İHA'ların sürü halinde hareket ettikleri görevler oldukça artmıştır. Birlikte gerçekleştirilen görevlerde hava araçları arasındaki koordinasyonun; havada olası çarpışmaların önlenmesi, birden fazla hava aracı tarafından toplanan ortak verilere göre kararlar alınması ve sürü içinde güvenli iletişimin sağlanması gibi temel gereksinimleri kapsaması gerekmektedir. Sürülerde kötü niyetli saldırılarla kaçırılma ve yok edilme tehlikesi yaşanabilmektedir. Hava aracının ele geçirilmesi hem sürünün güvenliğini hem de görevin sürdürülebilirliğini tehlikeye atabilir (García-Magariño et al., 2019). Ele geçirilmiş bir hava aracının sürüye göndereceği hatalı bilgilerin ayıklanabilmesi veya kötü niyetli kullanımların tespit edilebilmesi gerekmektedir.

Elektronik uçuş sistemleri bağlamında sayılamayacak olsalar da blok zincirinin havacılık endüstrisinin farklı ihtiyaçlarını karşılamak amacıyla geliştirilmiş yöntemleri de mevcuttur. Bunlar arasında dijital uçak bakım ve mürettebat sertifikaları, müşteri sadakat programları, bilet rezervasyon ve satış, kargo güvenliği gibi geniş bir yelpazede yer alan uygulamalar sayılabilir (Ahmad et al., 2021). Ancak bu bölümün kapsamı elektronik uçuş sistemleriyle sınırlandırıldığından aviyonik sistemler bağlamında yapılmış çalışmalar bu bölümde verilen alt başlıklara göre sınıflandırılarak Tablo 1'de verilebilir.

*Tablo 1. Blok zincirinin aviyonik sistemlere uygulandığı çalışmalar*

| Amaç | Geliştirilen Model/ Sistem | Çalışma |
|---|---|---|
| Uçuş Güvenliği | ATM, ADS-B, SWIM, BACS-IoD | (Arora & Yadav, 2019; Bonomo et al., 2018; Clementi et al., 2019; Reisman, 2019) |
| Milli Güvenlik Riskleri ve Mahremiyet | Teslimat konsepti, IoT | (Blank et al., 2018; Rahman et al., 2021) |
| Veri Güvenliği | FDR, LVBS, SEDIBLOFRA | (Barka et al., 2022; Bera et al., 2020; Joshi, Han, & Wang, 2018; Li, Jiang, Chen, Luo, & Wen, 2020) |
| Sürü Güvenliği | ABS, Ağ | (Bonomo et al., 2018; García-Magariño et al., 2019; Xiao et al., 2021) |

## 5. BLOK ZİNCİRİNİN UYGUNLUĞUNA KARAR VERİLMESİ

Hava araçları arasında kurulan blok zinciri ağındaki hava aracı sayısının artmasıyla hesaplama maliyeti, veri üretiminin artmasıyla da hafıza/disk alanı artar. Daha büyük veriler için özetleme (hashing) yapmak, daha fazla zaman, enerji ve işlem gücü gerektirir. Koordineli bir uçuş gerçekleştiriliyorsa geciken hesaplama veya iletişim hava araçlarının çarpışma riskini artırabileceğinden, blok zincir ağlarına yapılan siber saldırılarla blok zincirin güvenliğine tehdit oluşturulabilir. Burada bahsedilen olumsuz durumlar dışında blok zincir kullanımının, bazı ülkelerde hala regülasyonlara tabi olması ve bu nedenle, blok zincirin kullanımının bazen yasal olarak kısıtlanabilmesi de güncel bir sorun olarak görülebilir. Dolayısıyla bu teknolojinin belirli bir problemin çözümünde kullanılıp kullanılamayacağının veya hangi blok zincir türünün problem çözümünde daha elverişli olacağının da belirlenmesi gerekmektedir.

### 5.1. Blok Zinciri Kullanımının Sınırları

Blok zincirin aviyonik sistemlerde haberleşme, seyrüsefer, uçuş yönetimi, veri güvenliği gibi alanlarda kullanıldığı görülmektedir. Bunun yanında aviyonik sistemler, bu teknolojinin kullanılabileceği daha farklı alanları da barındırmaktadır. Özellikle sınır güvenliği, sürü içi haberleşme, hava trafik yönetimi, gözetleme, depo envanter kontrolü gibi alanlarda çok daha ilerleme potansiyeli mevcuttur. Ancak blok zincirin şeffaflık, güvenlik, verimlilik, değiştirilemezlik, taraflar arası güven sağlama gibi üstünlüklerinin yanında işlemlerin yavaş olmasından ve siber saldırılardan kaynaklanan eksiklikleri de bulunmaktadır.

Blok zincirinin eksikliklerinin başında özellikle herkese açık blok zincirin siber saldırılara maruz kalabilmeleri gelmektedir (Rodríguez-Molina, Corpas, Hirsch, & Castillejo, 2021). Performans açısından da blok zinciri, bloklar arasında yapılan işlemlerin sayısına ve boyutuna bağlı olarak yeterli seviyeleri yakalamayabilmektedir. Blok zincirin ölçeklenebilirliği, bloklar arasında yapılan işlemlerin sayısına ve işlemlerin boyutuna göre değişebilmektedir (Su, Wang, Xu, & Zhang, 2020). Bu nedenle de blok zincirin ölçeklenebilirliği ve performansı bazı durumlarda düşük olabilmektedir. Enerji tüketimi düşünüldüğünde ise blok zinciri teknolojisi, çalışırken yüksek miktarda elektrik tüketmektedir. Blok zincirin enerji tasarrufu genel olarak düşüktür (Zhenzhen, Xu, Fudong, & Jing, 2021). Öyle ki günümüzde blok zinciri kullanımı, bazı ülkelerde belirli düzenlemelere tabi tutulmuş, hatta yasal olarak kısıtlandığı yerler de olmuştur.

Blok zinciri siber saldırılara karşı da savunmasız kalabilmektedir. Zincirin güvenlik seviyesi, özetleme (hashing) yapma işlem gücüyle doğru orantılıdır. Zincir ağının bilgi işlem gücü arttırılarak saldırıların gerçekleştirilmesi daha zor hale getirilebilir. Bir blok zincirinin maruz kalabileceği saldırılar, %51, çift işlem, Finney, Vector76, kaba kuvvet, denge, Sybil, Netsplit ve Eclipse, madenci havuzu, DDoS gibi farklı şekillerde olabilmektedir (Oğuzhan & Kiani, 2018). Aviyonik sistemlerde uygulanan blok zinciri mimarileri de yapıları itibariyle benzer saldırılara maruz kalabilmektedir. Ancak sistemler, nesneler veya cihazlar arası ilişkiler bireysel kullanıcılarınkiyle aynı olmadığından saldırı türlerinin tehlike oluşturma kapasiteleri de farklılık göstermektedir. Bu saldırılardan aviyonik sistemlerde etkili olabilecekler ile bunlara karşı alınabilecek önlemler Tablo 2'de sunulmuştur.

*Tablo 2. Blok zincirine yapılan saldırı türleri*

| Saldırı Türü | Tanım | Önlem |
|---|---|---|
| Çift İşlem Saldırısı | Bir saldırganın, birbiriyle çelişen iki işlemi hızlı bir şekilde arka arkaya göndererek aynı anda çift işlem (örneğin finansal yapılarda çift harcama) yapma potansiyeline sahip olmasıdır. | Doğrulama sayısını artırma, ağa nöbetçiler yerleştirme |
| Finney Saldırısı | Kötü niyetli bir kullanıcının, henüz onaylanmamış bir işlemi diğer bir işlemin onaylanması anında ağa sokması ve ardından blok zincirinin bu onaysız blok üzerinden çatalını oluşturmasıdır. | Doğrulama sayısını artırma |
| Kaba Kuvvet (Brute Force) Saldırısı | Finney saldırısının geliştirilmişidir, saldırganın ağdaki çok daha fazla düğümü kontrol altına alarak çift işlem saldırısı için kullanmasıdır. | Ağa nöbetçiler yerleştirme |
| %51 Saldırısı | Bir madencinin blok zinciri hesaplama gücünün ve özet oranının %50'sinden fazlasını kontrol ederek blok zinciri ağını savunmasız hale getirmesidir. | Saldırganın boykot edilerek özetleme oranın %51 altına düşürülmesi |
| DDoS Saldırıları | Dağıtık Hizmet reddi saldırısı çok sayıda kaynaktan yapılan ve kullanıcıların hizmete erişimini sınırlamayı amaçlayan saldırıdır. | Merkeziliğin azaltılması |

Blok zincirin bu sınırlılıkları ile üstünlükleri onun aviyonik sistemlerde kullanılabilirliğini belirleyecektir. Bu sınırlılıklarıyla birlikte ihtiyaçları karşılayabilecek uygun bir yöntem olup olmadığına veya sınırlılıklar, üstesinden gelinebilecek düzeyde olduğu durumlarda çözümün blok zinciri teknolojisi yardımıyla gerçekleştirilmesine ve kullanılması gereken blok zinciri türüne karar verilmesi sıradaki adım olmaktadır.

## 5.2. Uygun Blok Zincirinin Seçilmesi

Aviyonik sistemlerde blok zincirinin kullanımı sayesinde güvenlik, değiştirilmez veri, anonimlik, merkezi olmayan veri yönetimi ve veriye erişim gibi özelliklerinden faydalanılabilmektedir. Aviyonik sistemler için blok zincirini kullanırken, öncelikle dikkat edilmesi gereken noktalar da vardır. Aviyonik sistemler genellikle hassas verileri işlediğinden bu verilerin güvenliği ve doğruluğu önemlidir. Blok zinciri teknolojisi, verilerin değiştirilmesini veya yanıltılmasını engel-leyerek güvenliği arttırabilir. Blok zinciri teknolojisi sayesinde, kaydedilen verilerin bir kısmı dahi olsa değiştirilemez. Bu da aviyonik sistemlerde önemli bir fayda sağlayabilir. Aviyonik sistemlerde verilerin kimin tarafından paylaşıldığının veya hangi kullanıcılar tarafından erişildiğinin gizlenmesi de önemlidir. Blok zinciri, kullanıcıların anonim olarak veri paylaşmasını sağlayabilir. Aviyonik sistemlerde, verilerin merkezi olmayan bir yapıda yönetilmesini blok zinciri sağlayabilir ve işletim maliyetlerini azaltabilir. Aviyonik sistemlerde, veriye erişimin kolay olması önemlidir. Blok zinciri teknolojisi ile oluşan veri tabanı tek merkezi değildir. Bu da aviyonik sistemlerde veriye erişimi kolaylaştırabilir (Li, Jiang, Chen, Luo, & Wen, 2017).

Diğer taraftan herhangi bir teknolojide olduğu gibi blok zincirinde de kullanım için bazı ticari, teknik veya hukuki kısıtlamalar olabilir. Ekonomik olarak blok zincirinde işlem maliyetleri yüksek olup aviyonik sistemler için uygun olmayabilir. Blok zincirinde veri saklama kapasitesi sınırlı olduğu için bazı aviyonik sistemler için yeterli olmayabilir. Güvenliği arttırmak için, blok zincirinin işlem hızının azaltılması gerekebilir. Ancak bu da blok zincirinin daha fazla işlemi doğrulamak için daha fazla zaman harcamasına neden olabilir. Performansı arttırmak için ise blok zincirin güvenliğinin azaltılması gerekebilir. Bu da blok zincirin daha az güvenli işlemleri kabul etmesine neden olabilir.

Blok zincirin merkeziyetsiz yapısını arttırmak için de veri depolama kapasitesi azaltılabilirken bu, sistemin daha az veri depolamasına neden olabilir. Güvenlik özelliğinden çok faydalanırken, işlem maliyetlerini azaltmak için merkezi olmayan veri yönetimi özelliğinden feragat etmek gerekebilir. Bu tür nedenlerle, aviyonik sistemler için blok zincirini kullanırken, faydalanılması düşünülen özellikler arasındaki dengeyi koruyan bir yol seçmek ve uygun olan özellikleri kullanmak gerekmektedir. Aviyonik sistemlerde blok zinciri kullanımına ihtiyaç olup olmadığı ile hangi blok zinciri türünün problem yapısına uygun olarak seçilebileceğine daha kolay karar verilmesini sağlayacak bir akış şeması Şekil 3'te verilebilir.

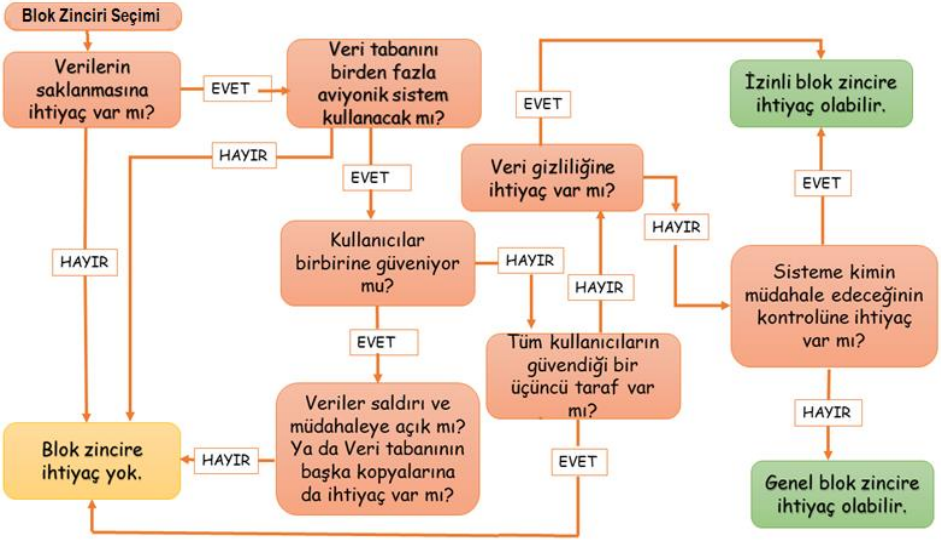

*Şekil 3. Uygun blok zinciri türünün seçilmesi adımları*

Genel olarak, aviyonik sistemlerde gerçekleşen her işlemin kaydedilmesi gerekmez. Örneğin bir aviyonik sistemde verilerin kaydedilmesi gerekmiyorsa blok zincir kullanmaya gerek yoktur. Nitekim kayıt ihtiyacı olmadan sadece iletişim gizliliği gerektiren sistemlerde bu ihtiyaç kriptolama yöntemleriyle de verimli şekilde giderilebilir. Diğer taraftan, verilerin saklanması gereken sistemlerde verilere ulaşan tek bir kullanıcı mevcutsa, blok zincir teknolojisini kullanmadan de veri güvenliği sağlanabilmektedir. Aviyonik sistemler arasında kullanılacak ve ortak işlemler yapılması gereken çok kullanıcının giriş yapabildiği sistemlerde gizlilik, güvenlik, şifreleme gibi ihtiyaçlar yoksa ve kullanılan klasik veri tabanları ihtiyacı karşılıyorsa blok zincir kullanımının getireceği faydalara yine ihtiyaç olmayabilir. Dolayısıyla böyle bir sistem için blok zinciri kullanımına yeterince ihtiyaç olmadığı savunulabilir.

## 6. BLOK ZİNCİRİNİN AVİYONİK SİSTEMLERDE KULLANILABİLECEĞİ YERLER

Hava aracı sayısının artmasıyla hava trafik güvenliğinin sağlanması da gün geçtikçe zorlaşmaktadır. Hava trafik yönetimi sistemlerinin gelişimini sınırlandıran ölçeklenebilirlik, dayanıklılık, tutarlılık ve gizlilik olmak üzere dört ana başlık mevcuttur. Hava trafik yönetim sistemleri açısından bir hava aracı sürüsü için yol planlaması yapan bir merkezi ağın tüm alanı kontrol edebilmesi oldukça zordur. Bu nedenle dağıtılmış ağlar önerilmektedir. Bu ağ merkeziyetsiz bir blok zincir ağı ile kontrol edilebilir. Genel anlamda bir uçuş planı, yetkili makamlara sunulmasından arşivlenmesine kadar birçok paydaşın erişim hakkına sahip olduğu uçuşla ilgili ayrıntıları içeren resmi bir işlemdir. Paydaşlar belgeyi güncelleyebilir ve belgeye veri ekleyebilirler. Paylaşılan verilerin bütünlüğü ve özgünlüğü, hava trafik yönetim hizmetlerinin hava trafiğindeki tıkanıklığı azaltma girişimlerinde kritik bir öneme sahiptir. Hava trafik yönetim teknolojilerinde iletişim, üçüncü kişiler tarafından dinlenme, iki birim arasındaki iletişiminin engellenmesi, mesajların silinmesi ve/veya değiştirilmesi gibi tehditlerle karşı karşıyadır. Uçuş planlama verimliliği ve güvenliğinde daha iyi performans ve daha gelişmiş güvenliğin anahtarı merkezilikten kurtulmaktadır. Blok zincirinin tek merkeze ihtiyaç duymayan mimarisi sayesinde hava sahasının her yerinde hizmetlerin ve verilerin kesintiye uğramadan kullanılabilmesi, uçuş planlarının değiştirilememesi ve kaydedilen bilgilerin sızdırılmaması sağlanır. Bu çözüm özellikle merkezi uçuş veri yönetimi mimarisinin de bulunmadığı bölgelerde hava trafik yönetim hizmetlerinin dağıtımını kolaylaştırabilir ("Automatic Dependent Surveillance-Broadcast (ADS-B),"). Blok zinciri teknolojisi, hava trafik yönetimi için uçuş verilerinin güvenli bir şekilde paylaşılması amacıyla kullanılabilir.

Uçuş kontrol ve seyrüsefer sistemlerinin yer kontrol istasyonundan gelen komutlara bağlı olduğu durumlarda ağdaki yoğunluk veya kopukluk tehlikeli sonuçlar doğurabilmektedir. Merkezi olmayan bir blok zinciri ağına sahip olmak, bu bağımlılığı ortadan kaldırabilir ve hava aracı ağlarının sinyal yoğunluğuna karşı savunmasızlığını azaltabilir. Her bir hava aracı, diğer hava araçlarının uçuş rotalarını içeren yerleşik bir blok zincir kopyasına sahip olur, bu sayede hedefine doğru ilerlerken diğer hava araçlarıyla olması gereken güvenli mesafeyi koruyabilir. Böylece hava araçlarının gecikmelere ve hatalara oldukça yatkın olan iletişim yoğunluğu azaltılabilir (Alladi et al., 2020; Mehta et al., 2020). Bundan kaynaklanan düşme, çarpışma gibi kazaların önüne de geçilebilir.

Blok zincirin getirdiği yeniliklerin başında verilerin güvenli bir şekilde saklanmasını sağlaması gelmektedir. Uçuş verilerinin kaydedilmesi ve saklanması amacıyla da blok zinciri kullanılabilir ki bunun için geliştirilmiş ve tescillenmiş bir yöntem mevcuttur (Croman et al., 2016). Dolayısıyla uçuş verileri blok zincir üzerinde kaydedilerek, verilerin doğruluğu ve güvenliği sağlanabilir. Bu sayede, uçuş verilerinin değiştirilmesi veya sahtecilik yapılması önlenebilir.

Geleneksel havacılık sistemleri, ticari anlaşmazlıkları çözerken çoğunlukla uzlaşma yöntemi olarak doğrudan müzakereyi veya merkezi arabuluculuk hizmetlerini tercih etmektedir. Yüksek maliyet ve uzun süreler, mevcut bürokratik uzlaşma yöntemlerinin önündeki önemli iki zorluktur. Akıllı sözleşme, katılımcı kuruluşlar arasında güven oluşturan, kendi kendini yürüten bir kod parçasıdır. Blok zincir teknolojisi, düğümler arası ağları kriptografik algoritmalarla birleştirerek, bir grup aracının belirli bir durum üzerinde bir anlaşmaya varabileceğini ve bu anlaşmayı bir kontrol otoritesine ihtiyaç duymadan güvenli ve doğrulanabilir bir şekilde kaydedebileceğini göstermektedir (Ferrer, 2016; Mehta et al., 2020). Bu yönüyle, blok-zincir teknolojisi kuruluşlar arasında iş birliğini kolaylaştırmak için güven ortamı oluşturmaktadır (Ahmad et al., 2021; Alladi et al., 2020; García-Magariño et al., 2019; Mehta et al., 2020; Rahman et al., 2021). Bu noktada da blok zinciri teknolojisi, akıllı sözleşmeleriyle güvenilirliği sağlar, iş kurallarını otomatikleştirebilir ve böylece anlaşmazlıkları aşabilir.

Nesnelerin interneti, izleme cihazları, sensörler, biyolojik yongalar gibi fiziksel nesnelerin birbirleriyle veya daha büyük sistemlerle veri bağlantısını ve paylaşımını sağlayan iletişim ağıdır. Nesnelerin interneti teknolojileri, tarım, lojistik hizmetleri, üretim süreci gibi birçok farklı sektöre uygulanabilmektedir. Nesnelerin interneti uygulamalarının etkin sahalarından biri de hava araçlarıdır, fakat hava araçlarının güvenlik saldırılarına karşı savunmasız olmaları, geniş kitlelerce kullanımlarını engellemektedir (Bera et al., 2020). Blok zincir teknolojisi kriptografik tekniklerin kullanımıyla güvenli, dağıtılmış bir depolama mekanizması sağlayarak, hava araçlarını nesnelerin interneti uygulamalarının vazgeçilmez bir parçası yapabilmektedir. Hava ve yer sensörü bileşenlerinin karşılıklı şekilde yararlı bir veri alışverişine katılabileceği, havadan yere endüstriyel nesnelerin interneti ağları için merkezi olmayan bir depolama mimarisi oluşturmak için blok zincir kullanılabilir (Rahman et al., 2021). Blok zinciri yardımıyla endüstriyel nesnelerin interneti ağlarını havadan yere kurmak oldukça yenilikçi bir çözümdür (Alladi et al., 2020; Bera et al., 2020). Fakat bu çözüm yöntemlerinin ileride etkin bir şekilde kullanılabilmeleri için farklı ticari işletmelerin işbirliği yapmaları gerekmektedir.

Blok zinciri, olası tehdit ve saldırılardan korunmak için açık anahtarlı kriptografi ve özel anahtarlı dijital imza şemaları kullanmaktadır. Tüm kullanıcıların açık anahtarları diğer tüm kullanıcılar tarafından bilinmekteyken, özel anahtarlar bireysel kullanıcılara özel olduğundan kimseyle paylaşılmamaktadır. Sistemdeki bir kullanıcı, o sistemdeki herhangi bir kullanıcıya veri göndermek istediğinde alıcının açık anahtarını kullanarak mesajını şifreler. Gönderilen bu veriyi alıcının özel anahtarı dışında hiçbir anahtar açamamaktadır (García-Magariño et al., 2019). Blok zincir, açık anahtar ve özel anahtar mekanizması yardımıyla bu tür ağlarda güvenli bire bir ve yayın iletişim olanakları sağlanmaya yardımcı olabilmektedir. Bu özellikten yararlanılarak geliştirilmiş karar verme mekanizmasındaki hava araçları arama ve tanımlama işlemi gerçekleşirken hava araçlarından biri hedefin tanımından emin değilse, onu ağda paylaşmakta ve diğer ağ üyeleri, hedefin kimliğine oy verebilmektedir. Oyların çoğunluğu hedefin gerçek kimliğini belirlemekte ve bu veriler ileride başvurmak üzere blok zincirinde saklanmaktadır (García-Magariño et al., 2019).

Blok zinciri teknolojisi, açık ve özel anahtar mekanizmalarıyla hava aracı ağlarında güvenli, bire bir iletişimi sağlamanın yanı sıra yayın iletişim olanakları sağlayarak da yardımcı olabilir ve en önemlisi, bu tip hava aracı ağlarında merkeziyetsizleştirilmiş veri depolayabilme imkanı sunabilir. Blok zincirinde her bir hava aracı, üzerine yazmadan kayıtları ekleyebilen bir düğüm görevi görmektedir. Depolanacak kayıtların kritik veriler olduğu durumlarda, dağıtık bir depolama hizmetine sahip blok zincir teknolojisi, ilgili hava araçlarının servis, bakım, sertifika gibi önemli kayıtlarını süresiz, şeffaf ve güvenli bir şekilde saklama imkanı sunabilir. Bu amaçla geliştirilmiş ve ticari amaçlarla kullanılan blok zinciri tabanlı güvenli kayıt yöntemleri de mevcuttur (Aleshi et al., 2019).

Uçuş planları da her bir hava aracı için kritik öneme sahiptir. Hava araçlarının özel, genel ve endüstriyel uçuşları sırasında kalabalık alanlarda uçmamasını sağlamak için gizliliğe duyarlı yol planlaması gereklidir. Bu aviyonik sistem ihtiyacını karşılamak üzere yapılmış çalışma da mevcuttur (Rahman et al., 2021). Özellikle İHA'lar için yol planlamasında uçuş sırasında gizlilik/mahremiyet gibi nedenlerle belirli bölgelerin üzerinden uçulmaması gerekebilir. Ayrıca planda irtifa, konum, rota, geçiş noktalarının sayısı, azami ve asgari hız sınırları gibi diğer kısıtlamalar da dikkate alınır. Gizliliğe duyarlı ve kısıtlamaya dayalı bir yol her zaman en uygun yol olmamaktadır. Hava araçlarının gizliliği ihlal etmeyi göze alarak maliyet açısından en uygun rota tercihinde bulunmamalarının kontrol edilmesi gerekebilir. Blok zinciriyle hava aracının belirlenen yol koordinatlarını kullanıp kullanmadığı kontrol edilebilir. Hava araçlarının uçuşu sırasında irtifası, hızı, yakıtı gibi mevcut durumuna yönelik bilgiler periyodik olarak toplanıp blok zincir ağına işlem olarak gönderilebilir. Blok zincir işlemin, daha önce belirlenen akıllı sözleşmelerle uyumu kontrol edilebilir. Blok zincirde saklanan bilgiler veri bütünlüğü, izlenebilirlik, değişmezlik gibi önemli ayrıcalıklar sağlayabilir.

Dağıtılmış karar verme algoritmaları, özellikle sürü hava aracı uçuşlarının geliştirilmesinde çok önemli bir rol oynayabilir. Dağıtılmış karar verme algoritmaları, dinamik görev tahsisi, toplu harita oluşturma ve engellerden kaçınma dahil olmak üzere birçok uygulamada kullanılabilir. Ancak fazla kullanıcının olduğu kolektif karar alma süreçlerinde karar hızı karar doğruluğu ile ters orantılıdır. Bu nedenle, dağıtık sistemlerde hava araçlarının karar vermesi konusunda daha otonom ve esnek çözümler gerekmektedir. Blok zincirinin sahip olduğu merkezi olmayan bir ağda ise tüm katılımcıların bir görüşü paylaşmaları ve oylamaları sağlanmaktadır. Hava araçları arasında bir konuda anlaşmaya varması için dağıtılmış oylama sistemi oluşturulabilmekte, böylece her bir hava aracı kendi oyunu şeffaf ve güvenli bir biçimde kullanabilmektedir (Ferrer, 2016).

Hava araçlarının afet yardımı, gözetleme, ağ aktarımı, enerji verimli cihaz keşfi gibi durumlarda da birbirleriyle koordineli hareket etmeleri gerekmektedir. Bu koordinasyonun sağlanabilmesi için havada çarpışmaları önlemek, birden fazla hava aracından toplanan verilere göre kararlar almak ve birbirleriyle güvenli bir şekilde iletişim kurmak gibi gereksinimlerin karşılanması gerekmektedir (Mehta et al., 2020). Hava aracı ağları savunmasız ve dinamik olduğu için düşman bir hava aracı ya da başlangıçta ağın bir parçası olan ancak daha sonra düşmanlar tarafından ele geçirilen hava aracı ağa girerek yanlış bilgi yayınlamaya başlayabilir veya orijinal verileri değiştirebilir. Böyle kötü niyetli girişimleri engellemek için blok zincirin fikir birliği mekanizması uygulanabilir. Doğal afet bölgelerinde yerel haberleşme sistemlerinin kullanım dışı kalması durumlarında örnek bir uygulaması (Joshi et al., 2018)'te verildiği gibi hava araçları için doğru, güvenilir bilgileri paylaşan bir blok zinciri destekli işbirlikçi hava-yer ağ mimarisi geliştirilebilirler. Kritik altyapıların koordineli siber fiziksel saldırılara karşı güvenliğini sağlamak için blok zinciri tabanlı bir hava araçları arası güven değerlendirme çözümü (Kim, KIM, & Lee, 2021)'te denendiği gibi geliştirilebilir. Ayrıca (Barka et al., 2022) tarafından sunulan uygulama, hava araçları tarafından toplanan hava durumu verilerinin güvenli ve etkili bir şekilde paylaşılması amacıyla blok zincirinin kullanılabileceğini göstermektedir. Blok zinciri, açık anahtar ve özel anahtar mekanizması yardımıyla bu tür hava aracı ağlarında güvenli bire bir ve yayın iletişim olanaklarının sağlanmasına yardımcı olabilmektedir. Aviyonik sistemlerde blok zinciri teknolojisinin uygulanabileceği alanlar özet olarak Tablo 3'te verilmiştir.

*Tablo 3. Blok zinciri uygulama alanları*

| Uygulanma Alanı | Üstünlükler | Eksiklikler |
|---|---|---|
| Uçuş Yönetim Sistemleri ve Uçuş Güvenliği | FMS ve haberleşme sistemleri arasında güvenli ve şeffaf bir veri paylaşımı sağlar. Uçuş planlarının değiştirilmesini ve kaydedilen bilgilere müdahale edilmesini engeller. Bakım kayıtlarının ve uçuş rotalarının blok zinciriyle saklanması uçuş güvenliğini artırır. | Blok zinciri teknolojisi daha karmaşık bir altyapı gerektirebilir ve veri depolama, işleme maliyetleri yüksek olabilir. |
| Milli Güvenlik Riskleri, Mahremiyet ve Veri Güvenliği | Verilerin güvenli ve değiştirilemez bir şekilde saklanmasını sağlar. Mahremiyet ve milli güvenlik risklerini azaltır, veri güvenliğini artırır. | Blok zinciri teknolojisinin entegrasyonu ve uyumu mevcut sistemlere bağlı olarak zor olabilir. Veri güncellemeleri ve erişimi için ek süreçler gerekebilir. |
| Haberleşme Sistemleri | Hava araçları arasında ve hava aracı ile yer istasyonu arasında güvenilir iletişim kanalı sağlar. Hizmetlerin ve verilerin kesintiye uğramadan kullanılabilmesini, bilgilerin sızdırılmamasını sağlar. | Blok zinciri kullanımı veri depolama gereksinimlerini artırabilir ve özel veri güvenlik önlemlerini gerektirebilir. Veri erişimi ve güncelleme süreçleri karmaşık olabilir. |
| Ticari Anlaşmalar | Hava aracı üreticileri, işleticileri, servisleri gibi şirketler arasında akıllı sözleşmeler yapılabilir. Bu anlaşmalar farklı bir kontrol otoritesine ihtiyaç duymadan güvenli ve doğrulanabilir bir şekilde denetlenebilir. | İleride etkin bir şekilde kullanılabilmeleri için farklı ticari işletmelerin kapsamla iş birlikleri yapmaları gerekmektedir. |
| Nesnelerin interneti | Güvenli, dağıtılmış bir depolama mekanizması sağlar. | İşlemlerin yüksek hızlarda yapılması zor olabilir. |

## 7. SONUÇ

Havacılık kuruluşları tarafından üretilen büyük miktarda veri, hızlı veri işleme gerektirmektedir. Blok zinciri teknolojisi üretilen bu verileri değiştirilemez, şeffaf ve güvenilir bir şekilde kaydedebilir. Akıllı sözleşmeleriyle güvenilirliği sağlayabilir. Taraflar arasında belirlenen prosedürleri otomatikleştirerek anlaşmazlıkları ortadan kaldırabilir. Böylece, katılımcı kuruluşlar arasında güven oluşturabilir. Hava aracı aviyonik sistemleri arasındaki veri akışının güvenliği kadar dış aviyonik sistemlerle olan bağlantılar da doğacak yeni ihtiyaçları karşılayacak şekilde yeni yöntemlerin uygulanmalarını zorunlu kılmaktadır. Nitekim farklı hava araçları ve sistemlerinin artışı bu sistemler arasındaki iletişimi saldırılara karşı daha büyük bir hedef haline getirmektedir.

Verilerin iletimi ve saklanması konusunda ise blok zincir teknolojisinin önemli bir potansiyele sahip olduğu görülmektedir. Yapılan çalışmalar, güvenli bir veri iletimi ve veri tabanı için blok zincirin önemli imkânlar sunduğunu göstermektedir. Bu çalışmaların başında blok zincir teknolojisinin sağladığı güvenilir iletişim imkânlarının kullanılarak hava araçlarının birlikte ve koordinasyon içerisinde hareket etmelerinin sağlanması ve böylece olası tehdit ve saldırıların üstesinden gelinmesi, çarpışmaların engellenmesi gösterilebilir. Ayrıca karar verme mekanizmalarında dağıtık bir yapı kullanma, yine farklı hava araçlarından gelen veriler sayesinde dağıtık ve çoklu veri toplama sayesinde iletişim güvenilirliği oldukça yüksek düzeylere çıkarılabilmektedir.

Aviyonik sistemler, çalışma şekilleri ve kapasiteleri itibariyle oldukça dinamik ve karmaşıktır. Akıllı sözleşmeler ve fikir birliği mekanizması gibi blok zinciri özelliklerini kullanarak otomatikleştirilebilir ve güvenli hale getirilebilir. Blok zincirin önemli bir avantajı olan dağıtık veri depolamaya sunduğu imkân kullanılarak hava ve yer verilerinin güvenilir bir şekilde depolandığı yöntemler geliştirilebilmektedir. Yine hava araçlarının kendi aralarındaki iletişimin güvenilirliği blok zincir kullanılarak sağlanabilmekte, elde edilen bilgilerin sadece kendi aralarında paylaşılmasına olanak sağlamaktadır.

Bu bölüm, blok zinciri teknolojisinin havacılık elektroniği açısından kullanılabilirliğini, çözüm önerilerini ve taşıdığı potansiyeli ortaya koymuştur. Blok zinciri teknolojisinin havacılık elektroniğinde sağlayabileceği temel fırsatlar tartışılmış ve özetlenmiştir. Havacılık sistemleri ve bileşenleri arasındaki temel hizmetleri ve etkileşimleri vurgulamak için blok zinciri tabanlı bir çerçeve sunulmuştur. Blok zinciri teknolojisinin başarılı bir şekilde kullanılmasını engelleyen çeşitli açık araştırma konuları ve zorluklar irdelenmiştir. Böylece oldukça yeni olan ve hızla gelişen blok zinciri teknolojisinin aviyonik sistemler açısından sunduğu fırsatlar ortaya çıkarılmıştır.

Bunun yanında bu alanda araştırma ve geliştirme yapmayı hedefleyen çalışanların blok zincirini hangi tür havacılık problemlerine uygulayabilecekleri veya fayda sağlaması zor olan konuları genel hatlarıyla belirlemektedir. Blok zincirin kendi problemlerinin çözümünde kullanılabilirliğini görebilmeleri için araştırmacılara yol haritalarını çizerken bir ön rehber sunulmuştur. Bu bölüm temel alınarak gelecekte aviyonik sistemlerdeki uçuş güvenliği, mahremiyet, veri ve sürü güvenliği konuları üzerine daha ileri ve kapsamlı çalışmalar yapılabilir. Uçuş yönetimi, haberleşme, hava trafik sistemleri için yeni yöntemler geliştirilebilir.

### TEŞEKKÜR

### KAYNAKLAR

Ahmad, R. W., Salah, K., Jayaraman, R., Hasan, H. R., Yaqoob, I., & Omar, M. (2021). The role of blockchain technology in aviation industry. IEEE Aerospace and Electronic Systems Magazine, 36(3), 4-15.

Akleylek, S., & Seyhan, K. (2018). Blok Zinciri Bileşenleri ve Uygulamaları Üzerine Bir Derleme. Paper presented at the İnformasiya təhlükəsizliyinin aktual multidissiplinar elmi-praktiki problemləri IV respublika konfransının materialları.

Aleshi, A., Seker, R., & Babiceanu, R. F. (2019). Blockchain model for enhancing aircraft maintenance records security. Paper presented at the 2019 IEEE International Symposium on Technologies for Homeland Security (HST).

Alladi, T., Chamola, V., Sahu, N., & Guizani, M. (2020). Applications of blockchain in unmanned aerial vehicles: A review. Vehicular Communications, 23. doi:10.1016/j.vehcom.2020.100249

Arora, A., & Yadav, S. K. (2019). Batman: Blockchain-based aircraft transmission mobile ad hoc network. Paper presented at the Proceedings of 2nd International Conference on Communication, Computing and Networking: ICCCN 2018, NITTTR Chandigarh, India.

Aslan, M. (2022). Blok zinciri ağ güvenliği ile konsensüs mekanizmaları ve akıllı sözleşmeler. (Yüksek Lisans). İstanbul Ticaret Üniversitesi, İstanbul.

Aslan, M., & Kasapbaşı, M. C. Blok Zinciri Platformları, Fikir Birliği Mekanizmaları ve Ağın Güvenlik Analizi. Haliç Üniversitesi Fen Bilimleri Dergisi, 5(1), 43-72.

Barka, E., Kerrache, C. A., Benkraouda, H., Shuaib, K., Ahmad, F., & Kurugollu, F. (2022). Towards a trusted unmanned aerial system using blockchain for the protection of critical infrastructure. Transactions on Emerging Telecommunications Technologies, 33(8), e3706.

Bera, B., Chattaraj, D., & Das, A. K. (2020). Designing secure blockchain-based access control scheme in IoT-enabled Internet of Drones deployment. Computer Communications, 153, 229-249. doi:10.1016/j.comcom.2020.02.011

Blank, P., Kirrane, S., & Spiekermann, S. (2018). Privacy-aware restricted areas for unmanned aerial systems. IEEE Security & Privacy, 16(2), 70-79.

Blasch, E., Xu, R., Chen, Y., Chen, G., & Shen, D. (2019). Blockchain methods for trusted avionics systems. Paper presented at the 2019 IEEE National Aerospace and Electronics Conference (NAECON).

Bonomo, I. S., Barbosa, I. R., Monteiro, L., Bassetto, C., de Barros Barreto, A., Borges, V. R., & Weigang, L. (2018). Development of swim registry for air traffic management with the blockchain support. Paper presented at the 2018 21st International conference on intelligent transportation systems (ITSC).

Chen, J., Du, C., & Han, P. (2016). Scheduling independent partitions in integrated modular avionics systems. PloS one, 11(12), e0168064.

Clementi, M. D., Larrieu, N., Lochin, E., Kaafar, M. A., & Asghar, H. (2019). When air traffic management meets blockchain technology: a blockchain-based concept for securing the sharing of flight data. Paper presented at the 2019 IEEE/AIAA 38th Digital Avionics Systems Conference (DASC).

Collinson, R. P. G. (2011). Avionics Systems Integration. In Introduction to Avionics Systems (pp. 459-487). Dordrecht: Springer Netherlands.

Croman, K., Decker, C., Eyal, I., Gencer, A. E., Juels, A., Kosba, A., . . . Wattenhofer, R. (2016). On Scaling Decentralized Blockchains, Berlin, Heidelberg.

Fedosov, E., Koverninsky, I., Kan, A., Volkov, V., & Solodelov, Y. (2017). Use of real-time operating systems in the integrated modular avionics. Procedia Computer Science, 103, 384-387.

Ferrer, E. C. (2016). The blockchain: a new framework for robotic swarm systems (2016). arXiv preprint arXiv:1608.00695.

García-Magariño, I., Lacuesta, R., Rajarajan, M., & Lloret, J. (2019). Security in networks of unmanned aerial vehicles for surveillance with an agent-based approach inspired by the principles of blockchain. Ad Hoc Networks, 86, 72-82. doi:10.1016/j.adhoc.2018.11.010

Gaska, T., Watkin, C., & Chen, Y. (2015). Integrated modular avionics-past, present, and future. IEEE Aerospace and Electronic Systems Magazine, 30(9), 12-23.

ISBN: 978-625-6971-84-4Güven, Ö. (2020). Dijital Dönüşümde Blokzincir Teknolojisi Ve Bitcoin'in Ekonomiye Etkisi. Aydin Adnan Menderes Üniversitesi Sosyal Bilimler Enstitüsü.

Han, P., Zhai, Z., & Zhang, L. (2020). A model-based approach to optimizing partition scheduling of integrated modular avionics systems. Electronics, 9(8), 1281.

Han, S., & Jin, H. W. (2014). Resource partitioning for integrated modular avionics: Comparative study of implementation alternatives. Software: Practice and Experience, 44(12), 1441-1466.

Higgins, F., Tomlinson, A., & Martin, K. M. (2009). Survey on security challenges for swarm robotics. Paper presented at the 2009 Fifth International Conference on Autonomic and Autonomous Systems.

Joshi, A. P., Han, M., & Wang, Y. (2018). A survey on security and privacy issues of blockchain technology. Mathematical foundations of computing, 1(2).

Juma, H., Shaalan, K., & Kamel, I. (2019). A Survey on Using Blockchain in Trade Supply Chain Solutions. IEEE Access, 7, 184115-184132. doi:10.1109/access.2019.2960542

Kamel Boulos, M. N., Wilson, J. T., & Clauson, K. A. (2018). Geospatial blockchain: promises, challenges, and scenarios in health and healthcare. In (Vol. 17, pp. 1-10): BioMed Central.

Kayton, M., & Fried, W. R. (1997). Avionics navigation systems: John Wiley & Sons.

Kim, K.-H., Kim, D., & Lee, S. (2021). United States Patent No. US20210354855A1.

Kumar, V. A. (2021). A Comprehensive Survey on Privacy-Security and Scalability Solutions for Block Chain Technology. Smart Intelligent Computing and Communication Technology, 38, 173.

Kuzmin, A., & Znak, E. (2018). Blockchain-base structures for a secure and operate network of semi-autonomous unmanned aerial vehicles. Paper presented at the 2018 IEEE International conference on service operations and logistics, and informatics (SOLI).

Li, X., Jiang, P., Chen, T., Luo, X., & Wen, Q. (2017). A Survey on the Security of Blockchain Systems. Future Generation Computer Systems, 107. doi:10.1016/j.future.2017.08.020

Li, X., Jiang, P., Chen, T., Luo, X., & Wen, Q. (2020). A survey on the security of blockchain systems. Future Generation Computer Systems, 107, 841-853.

Mehta, P., Gupta, R., & Tanwar, S. (2020). Blockchain envisioned UAV networks: Challenges, solutions, and comparisons. Computer Communications, 151, 518-538. doi:10.1016/j.comcom.2020.01.023
148

Millard, A. G., Timmis, J., & Winfield, A. F. (2013). Towards exogenous fault detection in swarm robotic systems. Paper presented at the Conference towards Autonomous Robotic Systems.

Miller, S. P., Cofer, D. D., Sha, L., Meseguer, J., & Al-Nayeem, A. (2009). Implementing logical synchrony in integrated modular avionics. Paper presented at the 2009 IEEE/AIAA 28th Digital Avionics Systems Conference.

Oğuzhan, T., & Kiani, F. (2018). Blok zinciri teknolojisine yapılan saldırılar üzerine bir inceleme. Bilişim Teknolojileri Dergisi, 11(4), 369-382.

Rahman, M. S., Khalil, I., & Atiquzzaman, M. (2021). Blockchain-Powered Policy Enforcement for Ensuring Flight Compliance in Drone-Based Service Systems. IEEE Network, 35(1), 116-123. doi:10.1109/mnet.011.2000219

Reisman, R. J. (2019). Air traffic management blockchain infrastructure for security, authentication, and privacy. Paper presented at the AIAA Scitech Forum.

Rodríguez-Molina, J., Corpas, B., Hirsch, C., & Castillejo, P. (2021). SEDIBLOFRA: A Blockchain-Based, Secure Framework for Remote Data Transfer in Unmanned Aerial Vehicles. IEEE Access, 9, 121385-121404.

Su, Z., Wang, Y., Xu, Q., & Zhang, N. (2020). LVBS: Lightweight vehicular blockchain for secure data sharing in disaster rescue. IEEE Transactions on dependable and secure computing.

Wang, H., & Niu, W. (2018). A review on key technologies of the distributed integrated modular avionics system. International Journal of Wireless Information Networks, 25, 358-369.

Xiao, W., Li, M., Alzahrani, B., Alotaibi, R., Barnawi, A., & Ai, Q. (2021). A blockchain-based secure crowd monitoring system using UAV swarm. IEEE Network, 35(1), 108-115.

Yaguchi, Y., & Wakazono, T. (2021). Flight Plan Management System for Unmanned Aircraft Vehicles Using Blockchain. Paper presented at the 2021 International Conference on Unmanned Aircraft Systems (ICUAS).

Zhenzhen, P., Xu, G., Fudong, Z., & Jing, L. (2021). A Blockchain-based Airplane Meteorological Data Sharing Incentive System. Paper presented at the 2021 IEEE 2nd International Conference on Information Technology, Big Data and Artificial Intelligence (ICIBA).

Zhu, Y., Zheng, G., & Wong, K.-K. (2019). Blockchain-empowered decentralized storage in air-to-ground industrial networks. IEEE Transactions on Industrial Informatics, 15(6), 3593-3601.